# The orbit and nature of the semi-detached stellar companion of Sgr A* super-massive blackhole.

Elia Leibowitz

School of Physics & Astronomy and Wise Observatory
Faculty of Exact Sciences
Tel Aviv University

eliamenl@gmail.com

Running title: **The semi-detached companion star of Sgr A* blackhole**

**Abstract**

In three previous papers I showed that the series of the midpoints of the times of all the X-ray flares of Sgr A* that have been detected so far harbor a statistical trend termed pacemaker regularity. This means that X-ray flares are detected more frequently around time points that are a subset of a periodic grid on the time axis of period $P_X = 0.1032\ day = 149\ minutes$. The series of the times of detection of the peaks of near IR (NIR) flares of the object are also regulated by a pacemaker with a period of $P_{IR} = 0.028\ day = 41\ minutes$. Here, I show that the series of the midpoints of the times of recorded NIR flares are also regulated by a pacemaker of the period $P_{IRM} = 0.039\ day = 56\ minutes$. The two pacemakers found in the previous papers were interpreted as signals of a star that revolves around the black hole of Sgr A* in orbit with a mean radius of ~3.2 Schwarzschild radii of the black hole, here corrected to ~3.13. The finding of the period of the third pacemaker is consistent with the suggested revolving star model. Here, I present the specific orbit of the star as well as a plausible description of its sidereal rotation. The model also implies that the star has an unusual internal structure. I show that the discovery of the GRAVITY collaboration of the motion of hotspots at distances from the black hole that are of the order of very few Schwarzschild radii of it may well be understood within the context of the revolving star model.

# 1. Introduction

This contribution is the fourth in a series of papers regarding the flaring activity of the X-ray and the near-IT (NIR) radiation source Sgr A* at the center of the Galaxy (Leibowitz, 2017, 2018, 2020, hereafter L1, L2, L3). They showed that the distribution of the times of all the recorded flares of the source in either of these two wavelength regions, since their first discovery some 19 years ago, have a statistical regularity termed modulation by a pacemaker. This means that the midpoints of the X-ray flares are detected favorably around time points that are part of a periodic grid of tick marks on the time axis of the period $P_X = 0.1032 \, day$. The frequency of the tick marks of the X-ray pacemaker $F_X = \frac{1}{P_X} = 9.6897 \, day^{-1}$ is not the frequency of occurrences of X-ray flares themselves. The value of $P_X$, as extracted from published observed data, was found statistically significant, i.e. not being a statistical noise fluctuation, at a $4.6\sigma$ level of statistical confidence.

The frequency of the pacemaker associated with the detection times of the peak intensity of recorded NIR flares was found to be $F_{IR} = 35.395 \, day^{-1}$. It was inferred from observed data at a $3.8\sigma$ level of statistical confidence that this frequency too is not a statistical noise fluctuation.

In an attempt to interpret the apparent non-random distributions on the time axis of these two sets of times I proposed in L3 a model of a star that revolves around the black hole (BH) of Sgr A* of the mass $\sim 4 \times 10^6 M(Sun)$. The BH is assumed to be non or slowly, rotating. This assumption has recently gained considerable supportive evidence by the work of Fragione and Loeb (2020). By analyzing the dynamics of members of the S star cluster that occupy the central parsec around the BH, confronted with the well-established lifetimes of these stars, Fragione & Loeb were able to establish a limit on the value of the parameter $\chi < 0.1$, where $\chi = 0$ corresponds to a nonrotating BH and $\chi = 1$ corresponds to rotation with the maximum allowed angular momentum value.

In the model proposed in L3, the orbit of the star is described as a precessing Keplerian ellipse with a semi-major axis of 3.2 R, where R is the Schwarzschild radius of the BH. This value is corrected in this paper to 3.1278 R (see section 4.3). The pacemaker frequency $F_X$ is identified as $F_{epi}$, the frequency of the epicyclic cycle of the star orbital motion around the center. This is the frequency of the periodically varying distance of the star from the BH due to some eccentricity of its quasi-Keplerian precessing orbit. The period $P_{IR} = \frac{1}{F_{IR}}$ is identified as $P_{orb}$, the mean period of the sidereal cycles of the binary revolution of the star around the BH. Here and elsewhere in this paper, unless stated otherwise, time is expressed in day units and frequencies in $day^{-1}$ units.

In Section 2 of this paper I report on a frequency of a third pacemaker that characterizes a set of times of IR flares that can be extracted from observed data. The value of this frequency, as not being due to statistical noise fluctuation, may also be estimated at a relatively high statistical confidence, although not as high as the confidence associated with the two other pacemakers in the system. In Section 3 I present an explicit expression and a plot of the star orbit in its orbital plane. Section 4 presents kinematic parameters of the star orbit as measured with a clock moving with the star, as well as with an identical clock in the hands of an observer on Earth. It also presents an equation that describes the position of the star along this orbit as a function of time. It is shown that the finding, at a relatively high probability, that the value of the frequency of the third pacemaker is not a statistical noise fluctuation is very much consistent with predictions that can be made on the basis of the model. Section 5 suggests that the star has an unusual internal structure. Section 6 is a discussion of the angular velocity of the star in its orbital motion. Section 7 is an attempt to understand, within the framework of the suggested revolving star model, the exquisite detection by the GRAVITY collaboration (GRAVITY Collaboration et al. 2018, hereafter GRAV18) of moving spots in the plane of the sky that are associated with NIR flares of the object. Section 8 is a summary of this presentation.

## 2. The $F_{IRM}$ Pacemaker

### 2.1. Data

Paper L3 presents a detailed statistical analysis of the set of times TIR269 and of its subset TIR255, the list of the times of the measured peak intensity of 269/255 NIR flares of Sgr A*. A major tool in that analysis is the frequency dispersion diagram (FDD), the detailed mathematics of which is presented in L2. Figure 2 in L3 presents the FDD of the TIR269 set of times in the frequency interval 0.8333<f<50. The deepest minimum in the FDD of TIR269 or of TIR255 in the range f>1.2 is around the frequency $F_{IR} = 35.395$. Statistical analysis applied on the TIR255 data set allowed us to accept at a $3.8\sigma$ level of statistical confidence that this particular frequency value at which the deepest minimum in the FDD is found is not due to statistical noise fluctuation

In L3 it was noted that when FDD is applied on a list of times of midpoints of NIR flares, rather than of times of peak intensity, the deepest minimum is still around $F_{IR}$ but the statistical significance that can be assigned to this result is considerable smaller.

Table 1 presents a list of times HJD-2450000 of midpoints of 240 flares of Sgr A* recorded between the years 2002 and 2017. I refer to it as the TIR240 set. These flares are subset of the flares referred to in set TIR255 of L3, for which the midpoints could be established.

| | | | | | |
|---|---|---|---|---|---|
| 2516.5621[1] | 2768.8212[1] | 2805.6614[1] | 2806.7712[1] | 2810.5180[2] | 3169.8617[14] |
| 3192.9170[9] | 3193.4820[7] | 3193.5337[7] | 3193.6480[7] | 3194.6427[3] | 3248.9397[4] |
| 3250.6978[4] | 3250.7399[4] | 3250.9808[4] | 3252.6901[4] | 3252.8196[4] | 3503.8545[7] |
| 3504.8514[7] | 3506.7564[7 | 3506.8192[7] | 3541.7434[7] | 3576.6527[7] | 3579.8071[8] |
| 3579.8667[8] | 3581.5587[7] | 3581.6260[6] | 3582.7527[7] | 3859.0153[8] | 3859.0469[8] |
| 3887.8004[14] | 3906.9093[8] | 3907.9090[8] | 3907.9503[8] | 3915.7460[7] | 3933.8028[8] |
| 4001.5264[7] | 4002.5167[7] | 4179.9065[7] | 4191.8136[10] | 4192.0518[10] | 4192.1237[10] |
| 4192.3842[10] | 4193.1352[10] | 4193.2585[10] | 4193.3919[10] | 4193.7308[10] | 4193.8357[7] |
| 4193.8857[7] | 4194.1232[10] | 4194.7430[10] | 4195.1233[10] | 4195.1900[10] | 4195.9345[10] |
| 4195.9886[10] | 4196.3231[10] | 4196.7307[10] | 4196.9887[10] | 4197.1224[10] | 4198.1197[10] |
| 4198.1868[10] | 4235.8098[6] | 4237.8259[7] | 4239.0023[8] | 4239.7343[7] | 4299.5350[13] |
| 4299.6118[13] | 4300.4711[13] | 4300.5485[13] | 4300.6975[13] | 4301.4878[13] | 4301.5519[13] |
| 4301.6724[13] | 4303.5432[13] | 4303.6706[13] | 4304.5039[13] | 4304.7431[13] | 4304.7695[1] |
| 4304.7966[13] | 4305.7403[7] | 4538.8705[7] | 4597.8287[7] | 4611.8311[14] | 4611.9190[14] |
| 4612.8253[11] | 4613.7621[14] | 4613.8381[14] | 4616.8781[14] | 4618.8862[14] | 4620.7359[5] |
| 4620.7876[5] | 4620.8439[5] | 4620.8998[5] | 4633.6045[7] | 4683.5565[7] | 4724.5321[7] |
| 4725.5785[7] | 4921.8976[7] | 4922.8331[12] | 4924.8558[12] | 4969.9145[14] | 5015.6810[7] |
| 5015.8004[7] | 5016.7744[7] | 5017.5730[7] | 5017.7702[7] | 5018.8292[7] | 5055.6140[7] |
| 5094.5070[7] | 5095.5211[7] | 5284.8585[7] | 5708.7402[14] | 5708.7944[14] | 6064.7245[14] |
| 6636.7473 | 6636.7788 | 6636.8487 | 6636.8914 | 6636.9347 | 6636.9566 |
| 6637.0412 | 6637.0979 | 6637.1576 | 6637.2226 | 6637.3021 | 6637.3996 |
| 6637.4588 | 6637.4921 | 6637.5187 | 6637.5768 | 6727.2320[15] | 6727.4021[15] |
| 6750.2116[15] | 6750.8329[15] | 6751.8781[15] | 6811.4721 | 6811.5145 | 6811.5684 |
| 6811.6756 | 6811.7458 | 6811.8377 | 6811.9007 | 6811.9196 | 6811.9402 |
| 6811.9597 | 6811.9906 | 6812.0639 | 6812.1072 | 6812.1361 | 6812.1911 |
| 6812.3015 | 6812.4147 | 6826.3137 | 6826.3543 | 6826.4173 | 6826.4817 |
| 6826.5116 | 6826.5767 | 6826.6446 | 6826.6808 | 6826.8074 | 6826.8574 |
| 6826.9020 | 6826.9998 | 6827.0481 | 6827.0862 | 6827.1030 | 6827.1319 |
| 6827.1666 | 6827.2095 | 6843.1483 | 6843.1896 | 6843.2489 | 6843.3863 |
| 6843.4822 | 6843.5141 | 6843.5523 | 6843.5854 | 6843.6501 | 6843.7125 |
| 6843.7637 | 6843.7970 | 6843.9034 | 6843.9809 | 6844.0349 | 7157.0463[16] |
| 7582.3112 | 7582.3401 | 7582.3652 | 7582.4567 | 7582.5127 | 7582.5867 |
| 7582.6332 | 7582.6623 | 7582.7459 | 7582.8071 | 7582.8498 | 7582.9088 |
| 7582.9323 | 7583.0237 | 7583.0837 | 7583.2155 | 7588.0351 | 7588.1387 |
| 7588.2712 | 7588.3233 | 7588.3599 | 7588.4087 | 7588.4475 | 7588.5275 |
| 7588.5873 | 7588.6466 | 7588.6929 | 7588.7495 | 7588.9348 | 7950.4978 |
| 7950.5716 | 7950.7157 | 7950.7928 | 7950.8425 | 7950.8851 | 7950.9123 |
| 7951.0759 | 7951.1920 | 7951.2649 | 7951.3117 | 7951.3319 | 7951.4277 |
| 7960.4932 | 7960.5925 | 7960.7204 | 7960.8913 | 7960.9705 | 7961.0569 |
| 7961.1453 | 7961.1895 | 7961.2625 | 7961.3335 | 7961.3912 | 7961.4447 |

Table 1: Set TIR240, the HJD-2450000 times of midpoints of 240 NIR flares of Sgr A* recorded between the years 2002 and 2017. Numbers in brackets refer to data sources as follows: [1] Genzel et al. (2003), [2] Eckart et al. (2004), [3] Eckart et al. (2006b), [4] Yusef-Zadeh et al. (2006), [5] Eckart et al. (2008b), [6] Eckart et al. (2008a), [7] Witzel et al. (2012), [8] Do et al. (2009), [9] Meyer et al. (2009), [10] Yusef-Zadeh et al. (2009), [11] Kunneriath et al. (2010), [12] Trap et al. (2011), [13] Haubois et al. (2012), [14] Shahzamanian et al. (2015), [15] Mossoux et al. (2016), [16] Fazio et al. (2016), no brackets: Witzel et al. (2018).

## 2.2. FDD of TIR240

Figure 1 is a plot of the FDD of TIR240, the set of midpoints of Sgr A* NIR flares. The frequency search interval is [1.2-65] and the number of sampled frequencies within this interval is $5.75 \times 10^6$. Explanations of the FDD search routine and of how the choice of the search parameter values is being made are given in L2 and L3. For the reason explained below, here the number of sampled frequencies is more than twice the suggested number according the considerations in L3. Arrow No. 1 in the figure marks the deepest minimum at the frequency 35.415 that is nearly identical to the frequency 35.3914 that was found already in L3 as the deepest minimum in the set of the IR flares midpoints. The reason for adapting the value $F_{IR} = 35.395$ as the frequency of the pacemaker of the set TIR255 is explained in L3. The second deepest minimum in Figure 1 marked by arrow No. 2 is at the frequency $F_{IRM} = 25.6888$. The third deepest minimum, marked by arrow No. 3 is at the frequency $F_I = 17.808$. Arrow 4 points at the seventh deepest minimum at $F_Z = 15.991$. The relevance of this frequency, as well as that of the other three that are marked by arrows in the figure will be explained in section 4.7.

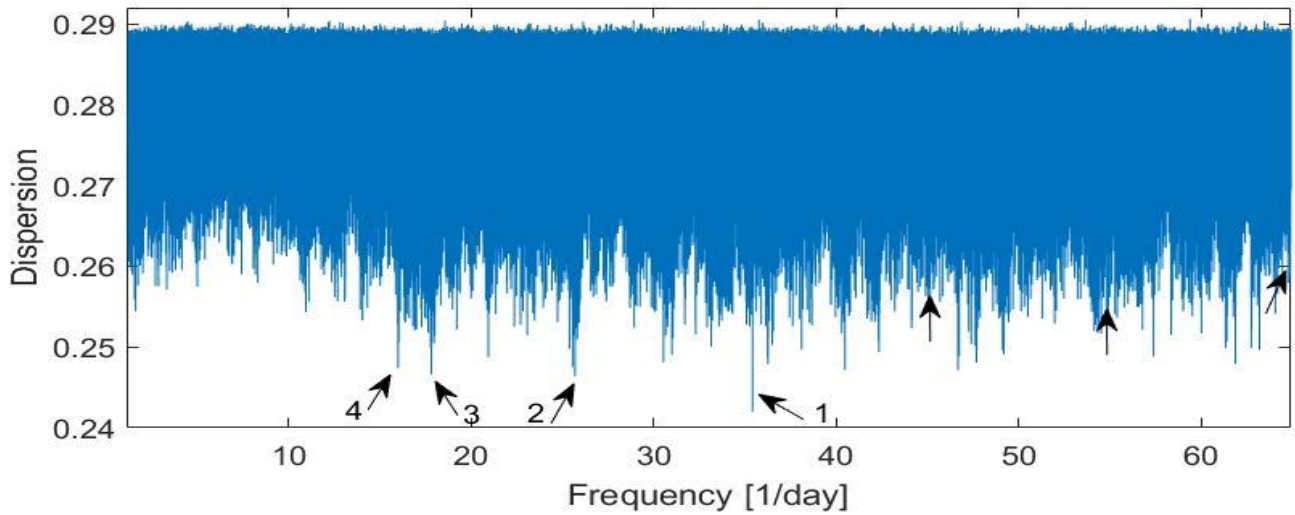

Figure 1: FDD of the data set TIR240, the times of the midpoints of 240 NIR flares of Sgr A* recorded between the years 2002 and 2017. Arrows 1, 2 and 3 point at the three deepest minima around the frequencies $F_{IR} = 35.415$, $F_{IRM} = 25.689$ and $F_I = 17.808$. Arrow 4 points to a minimum at the frequency $F_Z = 15.991$. See section 4.7 for the meaning of the frequencies marked by arrow 4 and the other three arrows in the figure.

## 2.3 Statistical significance

As explained in L2 and L3, the FDD is a plot of the parameter s, vs. all frequencies f within the frequency search interval. The dimensionless parameter s is the standard deviation (StD) of the distribution of the distances of the points of set TIR240, in the phase of the corresponding f frequency, from their corresponding nearest tick marks of the equi-distance grid on the time axis of the period p=1/f. Figure 1 seems like a rather noisy distribution of the s parameters over the frequency search interval. In particular, the two deepest minima are not qualitatively distinguishable from the next deepest ones seen in the figure. This is unlike the case of the FDD of sets TX83 and TIR255 analyzed in the previous papers of this series, in which a single outstanding minimum, at the frequency of the corresponding pacemaker, dominates the plot. In the case of TX83 there is also a second prominent dip in the FDD, although not as deep as the major one. This is also reflected in the numbers in Column 7 of Table 2. This table presents features of the three sets of times of recorded Sgr A* flares and their corresponding FDD function analyzed in this and in the three previous papers of this series.

| (1) Set | (2) event | (3) freq. Search interval | (4) No. tested freq's | (5) Dispersion | (6) Freq. deepest minimum | (7) prominence | (8) FNP |
|---|---|---|---|---|---|---|---|
| TX83 | X midpoints | 0.83-50 | $2.2 \times 10^6$ | 0.2083 | 9.6897 | 4.5 | $1/3*10^6$ |
| TIR255 | IR peaks | 1.2-50 | $2 \times 10^6$ | 0.2395 | 35.3914 | 7.9 | 1/8640 |
| TIR240 | IR midpoints | 1.2-65 | $5.75 \times 10^6$ | 0.2464 | 25.6888 | 3.7 | 1/967 |

Table 2: Characteristics of sets of detection times of three different definitions of flaring events of Sgr A* and their corresponding Frequency Dispersion Diagram. See text for explanations.

The columns in the table are (1) Name of the set that includes the number of elements in the set, (2) Flare time points defined as events the distribution of which on the time axis is being analyzed, (3) and (4) frequency search interval and number of equally spaced frequencies sampled within that interval on which the FDD is computed, (5) value of the statistic S, the StD of the observed time points dispersion around their corresponding nearest tick marks of the grid on the time axis of the period of the deepest minimum in the FDD. In the third line the reference is to the second deepest minimum in the diagram (see text below). The numbers in this column should be compared to 0.2887, the StD of a square probability density distribution over the [-0.5,0.5] interval on the number axis, (6) the frequency at which the FDD takes its minimum value S, (7) the distance of the S value obtained at the frequency of the FDD minimum point from the mean of the s parameter values at the frequencies of the 500 deepest minima in the diagram, in terms of the StD of these s values. This parameter is a measure of how prominent the S value is below the noise in the distribution of the minimum s values of the 500 deepest minima of the FDD. Note that the numbers in this column are presented only as some quantitative demonstration of the prominence of the signal of the pacemaker on the background of the apparent noise level in the plot. They are not utilized in the estimation of the statistical significance of the pacemaker found by the FDD. (8) The false negative probability (FNP) in negating the null hypothesis that the small value of S at the frequency of the deepest minimum is a random statistical fluctuation in the distribution of all the s values interrogated by the FDD.

The values of the FNP numbers in the two first lines of Column 8 in Table 2 were estimated in the previous papers, as explained in much detail in L3. In each one of these two cases, the estimate is based on the small value of the parameter S found by the FDD or on the s values at a few other minima that are close neighbors of the pacemaker frequency, the one that is the deepest in the FDD. It does not rely on any additional outside information. This is not the case here. Due to the apparent noisy background of the FDD plot as seen in Figure 1 and reflected in Column 7 of Table 2, the three statistical tests described in L3 yield a rather large FNP value in rejecting the null hypothesis that the value of S at frequency $F_{IRM}$ is due to statistical fluctuation rather than to some physical outside reality.

The value of the FNP presented in the bottom of Column 8 in Table 2 is estimated not on the small value of S at the $F_{IRM}$ frequency but rather on the fact that its frequency is the second deepest one in the FDD of the TIR240 set. This information is augmented by our a priori knowledge of a certain frequency value $F_D$ within the search interval of this FDD that we possess independently of set TIR240 that is under consideration. It is the difference between frequencies of two pacemakers that according to the null hypothesis are independent of one another and are also independent of the second pacemaker found in the TIR240 data det. We have that $F_{IRM} = 25.6888 \cong 25.7053 = 35.395 - 9.6897 = F_{IR} - F_X \equiv F_D$ .

Therefore, the frequency of the second deepest minimum in the FDD of TIR240 is not just a random number within the search interval. It is almost exactly the difference between two a priori known frequencies of pacemakers that are found to modulate the distributions of the times of Sgr A* flares. The difference between the frequency of the second deepest minimum in the FDD of T240 and the predetermined frequency value $F_D$ is $\delta = 25.7053 - 25.6888 = 0.0165$.

The probability that among the two frequencies of the two deepest minima in the FDD of TIR240 there will be, as a coincidence, at least one that is within the interval $[F_D - \delta, F_D + \delta]$ on the frequency axis, is: $FNP = 1 - \left(1 - \frac{2\delta}{63.8}\right)^2 = \frac{1}{967}$. Here, 63.8 is the width of the entire search interval. This FNP is the probability the we negate the null hypothesis that such a proximity of the two numbers is a statistical coincidence, while it is in fact an outcome of a purely random statistical fluctuation. The value of the FNP corresponds to a $3.28\sigma$ level of statistical confidence in the significance of the signal of the $F_{IRM}$ pacemaker in the observed data set TIR240. As said, the estimate of this statistical significance is based only on the fact that the dip at $F_{IRM}$ is the second deepest one in the FDD, and not on the value of the parameter S at this frequency. Therefore, the presence of a few other minima in the FDD of depths of a similar order of magnitude does not hamper this estimate.

The frequency $F_{IRM} = 25.7053$ can also be considered known a priory within the framework of the revolving star model suggested in L3. In it, $F_{IR}$ is identified as $F_{orb}$, the mean value of the frequencies of many cycles of the sidereal binary revolution of the star around the BH, and $F_X$ is identified as $F_{epi}$, the frequency of the epicyclic cycle in the orbital motion of the star (see Section 8 in L3). We therefore have also that $F_{IRM} \cong F_{orb} - F_{epi} = \gamma F_{orb} = F_{pre}$.
Here,

(1) $$\gamma = \frac{3GM}{c^2 a(1-e^2)}$$

is the general relativity (GR) precession angle of the quasi-Keplerian elliptical orbit around the BH per one cycle of the sidereal binary revolution. Here, a is the semi-major axis of the precessing elliptical orbit of eccentricity e, and $F_{pre}$ is the frequency of the GR precession (see Equation 3 and discussion in Section 8.1 in L3 and the following Section 3 of this paper). From this perspective, the finding of $F_{IRM}$ as the frequency of the second pacemaker in a set of times of flare events of Sgr A* can be viewed as a realization of a prediction made on the basis of the model. The FNP number is then the probability that this fulfillment of the prediction is only a matter of random coincidence.

The frequencies of the three deepest minima in the FDD of TIR240, marked by arrows 1, 2, and 3 in Figure 1 are very stable. FDD found precisely the same three numbers when it was computed on every one of all the tested $n_f$ numbers of frequencies within the search interval, from $1 \times 10^6$ to more than $5 \times 10^6$, and most probably also on much larger number of frequencies. The FDD presented in Figure 1 is computed on close to $6 \times 10^6$ sampled frequencies. There is however a slight dependence of the S values at these frequencies on the value of $n_f$. Figure 1 shows that the dips marked by arrows 2 and 3 have very similar depths. In FDDs computed on different $n_f$ numbers, for example, $[2 \text{ or } 4] \times 10^6$, the S value of the dip at $F_I = 17.808$ (arrow 3) is found to be slightly lower than the dip at $F_{IRM} = 25.6888$ (arrow 2), making the latter one the third rather than the second deepest minimum in the FDD. In that case, the exponent number 2 in the above FNP formula should be replaced by 3, giving the estimate $= \frac{1}{645}$. This corresponds to a $3.17\sigma$ level of statistical significance that can be attributed to the finding of the signal of the $F_{pre}$ pacemaker in the TIR240 data set. It should also be noted that $F_I = 17.808$ may be identified, within reasonable range of uncertainty, as a signal of the second harmonic of the deepest minimum in the FDD at the frequency 35.415. If this is the case, dip #3 in the figure is not independent of dip #1, and the exponent 2 in the FNP formula may perhaps be retained even if dip #3 is deeper than dip #2.

## 2.4. General comments regarding the FDD algorithm

The FDD algorithm has some technical resemblance to the well-known periodogram techniques constructed for searching periodicity in time series of irregularly spaced observations such as in Stellingwerf (1978) or Dworetsky (1983) papers. This may cause some misunderstanding regarding the nature of the FDD algorithm. In order to avoid it some further clarification may be warranted here.

There is a fundamental difference between the search of a periodic pacemaker in the distribution of elements of a single vector of points <t> on the time axis, as done here, and the search of periodicity in the two vector entity that usually makes the common time series of a light curve (LC), such as in the above-mentioned references. The first is in fact no more than a certain characterization of the distribution on the time axis of elements of the single vector of points <t>. It has gained only a little attention in the literature (e.g. Broadbent, 1955, 1956; Aschenbach 2010). Some statistical treatment of a single vector time series can be found also in the engineering literature (e.g. Steen and Lindsay 2015) but there it is more akin to the analysis in the pattern recognition research area. The vast majority of time series investigations in the astronomical literature are concerned with a two vectors entity of an LC type. The two vectors are of course the vector <t> of the times of the measurements and the vector <x> of the values of the measured dependent parameter, e.g. magnitude.

The claim of finding a periodicity in an LC is a claim that the value of the dependent parameter x measured at any specific time, in the past or in the future, should be within a certain interval of values, the width of which is determined by either estimated possible errors in the measurements or by uncertainty in the value of the claimed period. If at a certain time t, the x value is found to be outside this allowed uncertainty interval, the LC will be considered as quasi-periodic at best. The significance that is generally assigned to a suggested periodicity of a LC is the level of the statistical confidence that one may place in the belief that at any time t the measured value of the parameter x was, or will be in the future, within an interval of possible values as predicted by the suggested period.

In contrast, in finding a periodic pacemaker in the distribution of the elements of the single vector <t>, no claim is being made about the value of any measurable quantity that was not concretely measured. From its very definition it is clear that for a given finite interval on the frequency axis and a finite set of frequencies within it, every finite set of points <t> on the time axis has an associated pacemaker of a certain period P, or frequency F=1/P within that search interval. It is the frequency within the interval, with respect to the tick marks of its grid on the time axis, the points of <t> are grouped together most tightly, in the sense explained in Section 2 of L3. FDD is the algorithm that finds this P value, within the accuracy provided by the density of the distribution of the sampled frequencies within the search interval. This, in turn, must be defined, depending on the time interval spanned by the given set of times, so that all frequencies below the Nyquist one are properly interrogated. This has been explained in more detail in L2. The value of the statistic S mentioned above is quantifying the characterization of the distribution of the given set of points on the time axis. It also enables defining and ordering more than a single pacemaker of a set of time points, by the s value that corresponds to each dip in the FDD, when these are ordered by their depth in the plot, as was done in the previous section.

A periodogram of an LC consisting of a time vector <t>, and a corresponding dependent vector <x>, where all its elements are set as x=1, may reveal as the frequency of its highest peak a number that is close to the frequency discovered by the FDD algorithm applied on the one vector <t> time series. However, for a set of time points of finite number of elements and a given frequency search interval of finite width, the two search algorithms are not equivalent. This can be shown by examples of time sets for which the peak of the periodogram is around a frequency that is different from the frequency of the deepest minimum in the corresponding FDD.

This proves that neither one of the two methods is both necessary and sufficient condition for yielding the same results as obtained by the other. The reason is that they differ by optimizing different metrics applied to the distribution of the points: tightness of grouping around a set of evenly spaced points vs. residuals from a periodic model. Here, and in the previous papers of this series, we are concerned with the first one of these two different characterizations of the distribution of the points of set <t> on the time axis.

In order for the finding by FDD that P is the period of the pacemaker associated with a given vector <t> to be a matter of interest, one must show that there is only a small probability that the pacemaker has that P value just as a random coincidence. In other words, one must show that the FNP in rejecting the null hypothesis that the S value of the pacemaker found by the FDD is a random occurrence, is only a small number. For this purpose the results of the FDD process are put to statistical tests. Note, again, that FDD is exposing the period of the pacemaker associated with the set of times <t> and not a periodicity in the distribution on the time axis of the elements of <t> themselves.

Therefore, the findings that with respect to the frequency interval [1-50], $F_X = 9.6897$ is the frequency of the pacemaker associated with the set of the midpoints of the X-ray flares, and that within the interval [1.2-50] $F_{IR} = 35.395$ is the frequency of the pacemaker associated with the set of times of maxima of the IR flares, and that within the frequency interval [1.2-65] $F_{IRM} = 25.705$ is the frequency of the second pacemaker associated with the midpoints of the IR flares, are all well-established facts.

## 3. The orbit of the star

Equations (2) to (10) in Table 4 of L3 are post-Newtonian approximate expressions that determine the value of characteristics parameters of the orbit of the star around the BH. At that level of approximation we can write down the equation of the orbit itself in polar coordinates as

(2) $$r_t(\theta) = \frac{a_t(1-e^2)}{1+e\cos[(1-\gamma)\theta]}$$

Here, $r_t$ is the distance of the star from the BH and $\theta$ is the polar coordinate expressing an angle relative to some x-axis in the plane of the orbit, with the BH at the origin of the coordinate system as its apex. A state of the binary revolution when the star is at pericenter is taken as an initial condition $\theta_0 = 0$. The x-axis is in the direction of the line from the BH to the star at this state. From its initial $\theta_0$ value we consider the angle $\theta$ as increasing in its absolute value in the mathematical negative direction so that the radius-vector defined by it is rotating in the clockwise direction. This is done in order to conform with the direction of motion of the hotspots of the three NIR flares that were observed by the GRAVITY collaboration and are presented in their published maps (Grav18; GRAVITY Collaboration et al, 2020a, hereafter Grav20a). These maps will be referred to in Section 7. In equation (2) $a_t$ is the semi-major axis of the precessing elliptical quasi-Keplerian orbit of the star of eccentricity e . The parameter $\gamma$ is the GR precession angle, expressed as a fraction of 2π, per one cycle of the sidereal period of the star in its orbit around the BH of mass M. According to equation (3) in L3

(3) $$\gamma = \left(\frac{1}{2\pi}\right)\frac{6\pi GM}{c^2(a_t-R)(1-e^2)} = \frac{3}{2(a-1)g^2} .$$

Here, $a = \frac{a_t}{R}$ where $R = \frac{2GM}{c^2} = 2.9486 \times 10^{11} \times m \; cm$ is the Schwarzschild radius of the BH of mass M with $m \equiv \frac{M}{10^6 M(Sun)}$ and $g = \sqrt{1-e^2}$. By equation (3) in L3 we have that $\gamma = 1 - \frac{F_{epi}}{F_{orb}} = 0.7262$ is derived directly from the observations. With a known $\gamma$ value, equation 3 here expresses a dependence of a on the eccentricity e.

In the following discussion, all distances $r_t$ in the environment of the BH will be expressed in R units $r = \frac{r_t}{R}$ . Equation (2) can thus be written as

(2A) $$r(\theta) = \frac{ag^2}{1+e\cos[(1-\gamma)\theta]} .$$

The value of the parameter $\gamma$ can hardly be a rational number, therefore, equation (2A) represents an open orbit that is bound between an inner radius $r_{min} = a(1-e)$ and a maximum radius $r_{max} = a(1+e)$.

Figure 2 is a plot of equation (2A) for a few intervals of the parameter $\theta$ value, presenting sections of the orbit of the star, as seen with a line of sight normal to the orbital plane. The presented plots are for the choice of the parameters a = 3.1278 and e = 0.1712. These parameters of the orbit are uniquely implied by the identification of the observationally determined parameter values with the dynamical parameter values $F_X \leftrightarrow F_{epi}$ and $F_{IR} \leftrightarrow F_{orb}$ which stand at the core of the proposed revolving star model. The two measured parameter values imply also that m = 4.2184. All this will be explained in the following section.

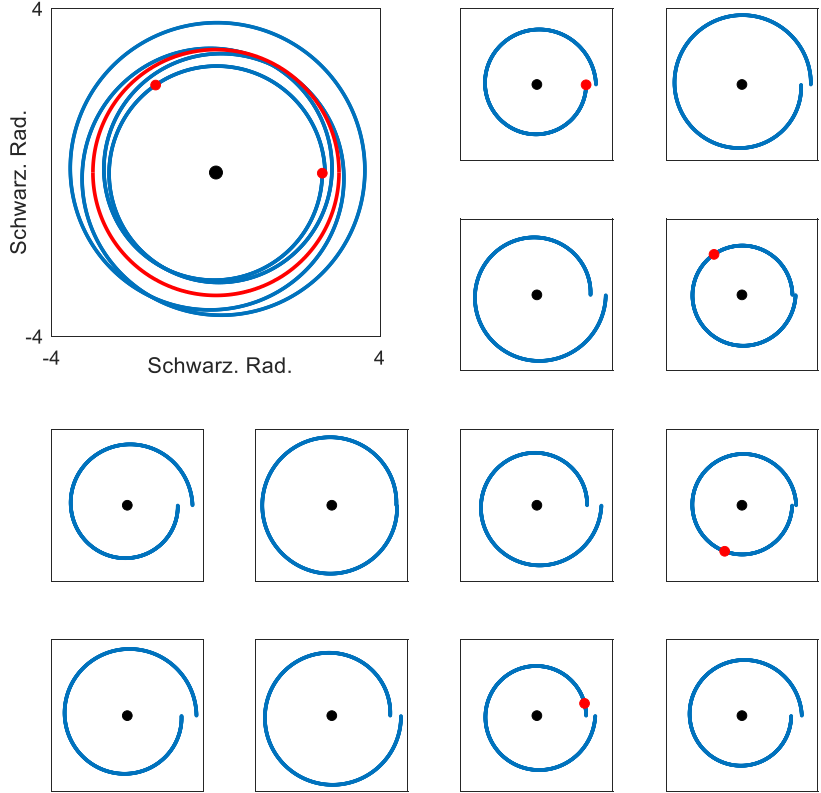

Figure 2: The large panel at the top-left corner depicts the orbit of the companion star of the BH of Sgr A*, covering four sidereal cycles of its binary revolution, measured from the position of the star at pericenter at polar coordinate $\theta_0 = 0$. The small panels are the first 12 sidereal cycles plotted individually. The x and y scales in all panels are [-4,4]. The black dot is the BH, red dots mark the position of the star at pericenters. The direction of rotation in all panels is clockwise. The red circle in the large panel is the innermost stable circular orbit of the BH. The parameters of the orbit are a=3.1278, e=0.1712, and $\gamma = 0.7262$.

The large panel at the top-left corner of Figure 2 depicts the continuous orbit of the first four successive sidereal orbital cycles of the star following the initial condition. These four cycles are seen individually in the 4 small panels to the right of the large one. The other eight panels in the figure depict the following eight sidereal orbital cycles of the star. The black dot is the BH and the red dots are the positions of the star on its orbit at pericenter passages. The extent of the frames in all panels of the figure is from -4 to +4 Schwarzschild radii with respect to the BH, in both the x and y directions. The direction of rotation is clockwise. The red circle is the ISCO, the innermost stable circular orbit around the BH.

## 4. Kinematics

### 4.1 Two clocks

The kinematical parameters of the star orbit can be presented as measured in two different time measurement systems, i.e. two clocks. The E clock is that of an observer on Earth, assumed to be at rest with respect to the BH. In the S system, time is measured with a clock that is identical to the E clock, attached to the center of the revolving star. Parameter values as measured with the S clock will be denoted with an apostrophe. Parameters that have the same numerical values in the 2 systems will be denoted as measured in the E frame.

The transformation from the S to the E clock is as follows. If $\Delta t'$ is the time interval between two events as measured with the S clock, then the time interval between these two events as measured in the E system is

(4) $$\Delta t = \frac{\Delta t'}{\beta(r)},$$

where $\beta(r)$ is a factor expressing the GR and the special relativity (SR) time dilations between the E system and the S system when the S clock is on the quasi-Keplerian orbit of the star at a distance r from the BH. According to Equations (5) and (6) in L3

(5) $$\beta(r) = \sqrt{\frac{(2r^2-5r+2)(r-2)}{2(r-1)^3}}$$

### 4.2. Time of passage of the star through orbital points

I divide a 2π cycle of the angle θ into $N_c = 1000$ small equal increments $\Delta\theta = \frac{2\pi}{N_c}$. For each integer k there corresponds a $\theta_k = k \times \Delta\theta = \frac{2\pi k}{N_c}$ value, which is a polar coordinate, relative to $\theta_0 = 0$, the polar coordinate at the star initial state. To each $\theta_k$ value equation (2) is matching an $r_k$ value such that the pair $[r_k, \theta_k]$ defines a point on the orbit. Using Kepler's second law, to the increment $\Delta\theta$ when the star is at distance $r_k$ we can associate the time difference

(6) $$\Delta t'_k = t'(\theta_{k+1}) - t'(\theta_k) = \frac{P'_{orb}\sqrt{1-e^2}}{N_c}\left[\frac{1}{1+e\cos[(1-\gamma)\theta_k]}\right]^2$$

It is apparent in Figure 2, that the distance r of the star from the BH is varying considerably as the star moves along its orbit. So does also the orbital period of the sidereal cycle of the revolution (see section 4.5). Therefore, the parameter $P'_{orb}$ on the right-hand side of equation (6) must be understood as the period of the mean of many sidereal orbital cycles of the star.

The total time elapsed from $t'_0 = 0$, to the time of the star passage through the $[r_n, \theta_n]$ point, is

(7) $$t'_n = \sum_{k=1}^{n} \Delta t'_k$$

Equations (6) and (7) are valid in the S system. In particular, $P'_{orb}$ is a parameter that is not measurable directly. However, under the assumption $F_{IR} \leftrightarrow F_{orb}$ we know that $P_{orb} = \frac{1}{F_{IR}}$, and $P_{orb}$ is the corresponding mean period in the E system. By equations (4) and (5) we have that

(8) $$P'_{orb} = \beta(\eta a) P_{orb} \ .$$

The parameter $\eta$ is introduced here in order to express the (still unknown) semi-major axis of the ellipse of the sidereal cycle with the mean period $P'_{orb}$. The parameter $\eta$ is presenting this unknown semi-major axis as a fraction of a, the value that according to equation (3) corresponds to the e value under consideration.

According to equation (4), the time interval between the two events, the star at the polar coordinate $\theta_k$ and the star at $\theta_{k+1} = \theta_k + \Delta\theta$ is measured in the E system as

(9) $$\Delta t_k = \frac{\Delta t'_k}{\beta(r_k)} \ ,$$

And the time of the passage of the star through the point $[r_n, \theta_n]$ is

(10) $$t_n = \sum_{k=1}^{n} \Delta t_k \ .$$

## 4.3. Calculated orbital frequency

We can now determine directly the period of each individual cycle of the sidereal revolution of the star, such as the one measured from a certain time $t_j$ following the initial state

$$(11) \qquad P_{orb}^{j} = t_{j+N_c} - t_j$$

A calculated mean orbital period will then be $P_{orb}^{cal} = \frac{\sum_{j=1}^{N_o} P_{orb}^{j}}{N_o}$ where $N_o$ is the number of sidereal cycles for which the calculations (by equation (2A)) is performed. Here $N_o = 1000$.

For a given $F_X$ and $F_{IR}$ the value of the calculated frequency $F_{orb}^{cal} = \frac{1}{P_{orb}^{cal}}$ depends on the two input parameters $e$ and $\eta$. I consider the parameter $\delta P = |F_{IR} - F_{orb}^{cal}|$ and scan over all points in the $[e, \eta]$ plane, for $0.15 \leq e \leq 0.4$ and $0.9 \leq \eta \leq 0.99$, calculating the value of $\delta P$ corresponding to each $[e, \eta]$ pair. I find a minimum of the $\delta P$ parameter at $e_a = 0.17121$ and $\eta_a = 0.97069 = 1 - e_a^2$, with $\delta P = 1.2257 \times 10^{-6}$. The semi-major axis of the precessing ellipse corresponding to this $e_a$ value is determined by equation (3). Together with the $\eta_a$ value, the mass of the BH is then given by

$$(12) \qquad m = \beta(\eta_a a)\left(\frac{1}{\sqrt{a}(a-1)}\right) \times 989.35 \times P_{orb}$$

Here, $\qquad 989.35 = \frac{c^3}{\sqrt{32\pi}G}\left(\frac{24 \times 3600}{10^6 \times 1.988 \times 10^{33}}\right)$.

These implied $a$ and $m$ values are the parameters of the orbit displayed in Figure 2. Note that by adopting the revolving star model and using the Paczynsky-Wiita (1980) approximation for the SR and GR equations in the spacetime of the Schwarzschild metric, once the values of the observables $F_X$ and $F_{IR}$ are known, there is no free parameter in the derivation of the BH mass m, and of the a and e parameters of the revolving star orbit. Considering the uncertainty of ±0.03 in the value of $F_X$ and of ±0.5 in the value of $F_{IR}$, as estimated in L3, we find the parameters of the Sgr A* binary system within the following uncertainty limits:

$$m = 4.2184^{+4.2299}_{-4.2011} \qquad a = 3.1278^{+3.1607}_{-3.1012} \qquad e = 0.1712^{+0.1943}_{-0.1526}$$

## 4.4 The epicyclic frequency

Using Equations (2A) with $N_c = 1000$, I computed the value of r, the distance of the star from the center, along $N_o = 1000$ sidereal orbital cycles of the system, lasting 28.26 days as measured by the E clock. With Equation (7) and (10) I computed the $10^6$ values of t' and of t, the corresponding time points in the S and the E systems. In Figure 3a1, on the left-hand side, is a plot of r(t'), the distance of the star from the center as measured by an observer on the S system. Panel a2 on the right-hand side is the distance r(t) as measured by an E observer equipped with an identical clock. The figures display this function along only half a day following the initial condition. As can be seen, r varies cyclically between the values $2.592 = a(1-e)$ and $3.663 = a(1+e)$. Figure 3a1 displays 8.59 cycles of the period $P'_{epi} = 0.05821\, day = 83\, min$ which is the epicyclic period in the S system. Figure 3a2 on the right shows that in half a day the E observer detects 4.85 cycles of the period $P_{epi} = 0.1032\, day = 148.6\, minutes = P_X$. The mean distance of the star from the BH over the entire range of the computations is $mean(r) = 3.0819 \cong 3.0816 = a\sqrt{1-e^2} = b$, where b is the semi-minor axis of the precessing ellipse.

As explained in L3, according to our model, X-ray flares occur when the Roche lobe of the star has its smallest radius, i.e. when the star is near pericenter. Events of intense mass loss from the star, which are the energy sources of the X-ray flares, do not occur every time the star is near pericenter. They do take place, on the average, once every nine or 10 such passages, when the intense tidal waves in the outer layers of the star, that are excited by the oscillating distance of the star from the BH, exceed a certain threshold amplitude.

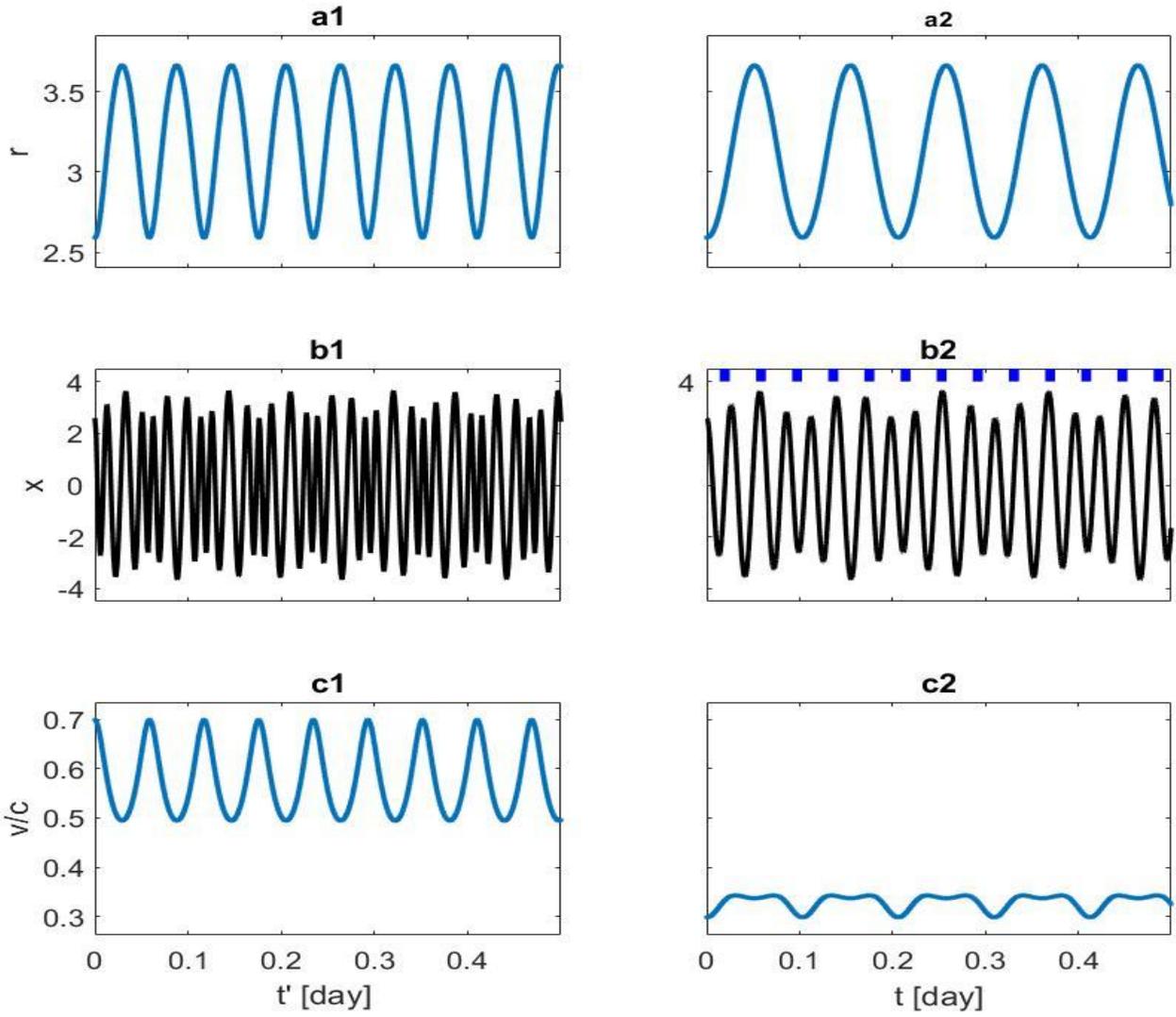

Figure 3: (a1) The distance of the star from the BH as a function of time over half a day, as recoded in the S frame. (a2) The same in the E frame. (b1),(b2) The value of the x coordinate of the position of the star in the orbital plane along half a day in the two time measurement systems. Tick marks on the upper frame line of panel (b2) constitute a grid with the period $P_{pre} = \frac{1}{F_{pre}} = \frac{1}{25.69}$. (c1),(c2) The velocity $\frac{v}{c}$ of the star as measured with the S and the E clocks.

The response of the outer layers of the star to the huge tidal force exerted on them by the BH, which varies with a very large amplitude at the $F_{epi}$ frequency, is very likely a major heat source in the internal energy budget of the star. The varying deformation of the outer envelope of the star may have a profound impact also on the thermodynamic and magneto-hydrodynamic characteristic of these layers and they may well be a major source of very large magnetic and flaring activity of the star. One expects that the entire internal structure of the star will be quite different from that of stars of various luminosity types that are common in the Galaxy. This will be further discussed in Section 5. However, trying to build a model for this star under these extreme circumstances is beyond the scope of this paper.

## 4.5 The sidereal orbital frequency

Figure 3(b2) presents a half day section of the time series of the coordinate x of the revolving star as measured in the E system. In panel (b1) on the left-hand side it is in the S system. As above, it was computed for 1000 orbital cycles. As seen in the figure, during the half day presented in panel (b2), the x coordinate made 17.7 oscillations around the x=0 value. In other words, the star made that number of sidereal orbital revolutions. As could be seen in the figure, the period of the sidereal orbital revolution is varying in time between 0.0261 and 0.0319 at the $F_{epi}$ frequency. The mean period over its 1000 revolutions considered is $P_{orb} = 0.028253$. The corresponding mean frequency is 35.394998, as noted in section 4.3. The varying value of the sidereal binary frequency around this mean is a major cause of the dispersion in the grouping of the NIR flares peak times around the tick marks on the time axis of this pacemaker (see section 8.3.2 in L3).

## 4.6 The velocity of the star in its orbit

The instantaneous velocity of the star at each point on its orbit is $v = r\frac{d\theta}{dt}$ where t here is either t of the E frame or t' of the S frame. In our numerical computation, at each point $[r_n, \theta_n]$ on the orbit, in the S system we have $v'_n = r_n \frac{\Delta\theta}{\Delta t'_n}$ and in the E system $v_n = r_n \frac{\Delta\theta}{\Delta t_n}$. Figure 3 (c1) and (c2) depict the velocity $\frac{v}{c}$ of the star (relative to the BH), along half a day, as recorded in the two clocks. The oscillations of the velocity seen in panel (c1) with the frequency $F'_{epi} = 17.174$ are due to Kepler's second law. The peculiar shape of the curve in panel c2 that oscillates with the prime frequency $F_{epi} = 9.6897$ is due to the superposition, over the Kepler's velocity variations, of the variations that are due to the varying GR and SR dilation factors, as the star is cyclically approaching and receding from the BH, and thus affecting dramatically the measurements performed on Earth. The mean velocities of the star are $\langle\frac{v'}{c}\rangle = 0.5969$ and $\langle\frac{v}{c}\rangle = 03283$ .

## 4.7 The sidereal precession frequency

Figure 4 is the power spectrum (PS) of the entire computed x(t) function. It is dominated by six high peaks. The three highest peaks in the figure are truncated in order to enable seeing the three lower ones that are not seen by eye in a full scale figure. A synthetic time series consisting of a combination of six Sine waves with these 6 frequencies is indistinguishable from the curve seen in figure 3(b2).

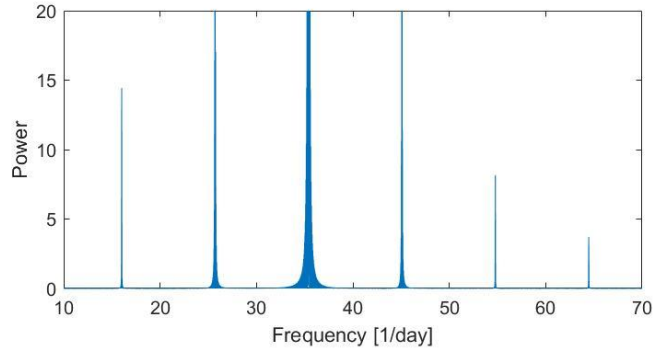

Figure 4: Power spectrum of the function x(t) of panel (b2) in Figure 3.

The highest peak near the center is around the frequency $F_{orb} = 35.399$. This is nearly the exact frequency of the deepest minimum in Figure 1 marked by arrow No. 1. The two peaks on its left are at the beat frequencies $[25.708, 16.016] = F_{orb} - k \times F_{epi}$, $k = 1:2$. They are marked in Figure 1 by arrows No. 2 and 4. The three peaks on the right are at the frequencies $[45.089, 54.779, 64.470] = F_{orb} + j \times F_{epi}$, $j = 1:3$. They are marked by the three other arrows in figure 1.

For an orbital plane that is inclined by an angle $i \neq 0$ with respect to the plane of the sky, where the y-axis at x=0 is the line of nodes, when x>0 (say) the star is slightly in front of the BH relative to the observer and at x<0 the star is slightly behind. The changing aspect ratio of the star and the BH with respect to the line of sight from Earth due to the binary revolution of the star is translated into periodic variations in the physical conditions that affect the transfer of NIR radiation emanating from the close vicinity of the star in the direction of Earth. Hamaus et al (2009) showed that in the observed light curve of IR radiation of a source that revolves around a BH, when the inclination angle of the orbital plane is $i \neq 0$, there is a periodic variation in the recorded brightness of the source at the sidereal orbital frequency. In L3 it was shown how this effect explains well the origin of the $F_{orb}$ frequency of the pacemaker that was found to modulate the recorded times of the peak intensity of NIR flares. The agent responsible for this periodic pacemaker is the harmonic component of the x(t) time series that has the frequency $F_{orb} = 35.395$.

The time series x(t) shows a clear signal of its harmonic component of the physical meaningful frequency $F_{orb} - F_{epi} = F_{pre}$. This is the frequency of the harmonic component of x(t) with the second largest amplitude, as mentioned above. The tick marks in the upper frame line in panel (b2) mark the positions on the time axis of the points of maxima of this harmonic wave. They constitute a grid with the $F_{pre}$ frequency. For an inclined orbital plane, the figure shows that while the star is changing its position from being in front of the BH to being behind it at the orbital frequency $F_{orb}$, it is also spending half of its time in front of the BH and one-half behind it, alternating between these two positions at the $F_{pre}$ frequency as well. Due to the varying radiation transfer circumstances mentioned above, there is half a cycle of the $P_{pre} = \frac{1}{F_{pre}}$ period during which the intensity of the NIR radiation originated at close vicinity of the star is measured by an observer on Earth statistically fainter than the intensity emitted, while the star is dwelling in the other half of this cycle. This creates an imbalance between the number of detected events that occur during one-half of this period and the number of detected events that occur in the other. This explains well the discovery of the third pacemaker with the $F_{pre}$ frequency that modulates the midpoint times of recorded NIR flares of Sgr A*.

In L3 the FDD was applied on set IR255, the list of times of the peak intensity in 255 NIR flares. The deepest minimum in the FDD of that set is at the frequency $F_{orb}$, which was found to be of high statistical significance. No signal of an $F_{pre}$ frequency was found in that list of times.

For set IR240, figure 1 shows that the deepest minimum in the FDD is still at $F_{orb}$, although, as noted already in L3, it is a less pronounced feature than in the FDD of set IR255. The second deepest minimum is at $F_{pre}$, as discussed in section 2.

In set IR240, the times that are considered the occurrence times of an NIR flare are the midpoints of the flares on the time axis. They differ from the flare peak times, and have a different distribution on the time axis than that of the points of the IR255 set. This difference is reflected in the different values of the corresponding two pacemaker periods $P_{IR} \leftrightarrow P_{orb}$ and $P_{IRM} \leftrightarrow P_{pre}$ derived from the observed data. It is much consistent with consequences of the revolving star model as exhibited in frame (b2).

We have that $\frac{P_{pre}}{P_{orb}} \cong 1.38$ and that the two periods are non-commensurate. Over a few $P_{pre}$ cycles, the times of events that occur at a given phase in the $P_{orb}$ cycle are distributed nearly evenly along the phase of the $P_{pre}$ cycle. Therefore, the times of the peak intensity of flares do not show the $P_{pre}$ pacemaker signal in their distribution on the time axis. Accordingly, no outstanding minimum is found around $F_{pre}$ in the FDD of the IR255 set. On the other hand, the midpoints of flares are still grouped favorably around the tick marks of the $P_{orb}$ pacemaker since a cycle of the $P_{pre}$ period includes one, and only one, full cycle of the $P_{orb}$ periodicity. Therefore, the FDD is uncovering the $P_{orb}$ pacemaker also in the distribution of the IR240 set of times.

Note that in the six sine wave presentation of the x(t) time series, the two components with the largest amplitudes are those with the frequencies $F_{IR}$ and $F_{IRM}$. These are frequencies of processes in the real physical world, namely, the sidereal and precession cycles of the orbit of the BH of Sgr A* companion star. These are also the frequencies that the FDD analysis of the set TIR240 uncovered most clearly in the observed data. The harmonic component of the x(t) function, represented in Figure 4 by the extreme left peak at the frequency 16.016 can be identified in the FDD of TIR240 as the fourth deepest minimum in figure 1, at the frequency $F_Z = 15.991$, marked by arrow No. 3. The frequencies of the other three harmonic components of x(t), marked in Figure 1 by the three arrows to the right of the center are practically indistinguishable within the noise apparent in the figure. These four frequencies can hardly be directly associated with any simple physical process in the Sgr A* system. Accordingly, as should be expected, there is only weak, or practically no real signal of them, that can be found in the observed data.

## 5. The internal structure of the star

For a star of mass $m_S$ at distance $a_S$ from a large mass M, equation (2) of Eggleton (1983) presents the radius of the Roche lobe of the star as a function of $a_S$ and the mass ratio $q = \frac{m_S}{M}$. In our model, at pericenters of the star orbit, when the distance of the star from the BH is $d_{peri} = a(1-e) = 2.5923$, the star is filling its Roche lobe. With m=4.2184, Eggleton equation describes a mass-radius (M-R) relation for that star. The cyan colored curve in Figure 5 depicts, on a logarithmic scale, this M-R relation. The upper red curve in the figure is an empirical M-R relation of Galactic low mass stars, as presented by Demory et al (2009). The blue curve at the bottom is a representative M-R relation of white dwarfs (WDs), taken from Figure 13 of Romero et al (2019).

The Eggleton formula is strictly valid for a star that rotates in synchronization with its orbital angular velocity. This may be nearly the case here as will be discussed in the next section. However, following Rees (1988) I have also obtained a direct expression for $r_S$, the distance from the center along the inter-binary axis where in the frame of reference that rotates with the orbital angular velocity of the star, an equilibrium between the gravitation and the centrifugal forces is achieved.

$$(13) \qquad r_S = \left(\frac{m_S}{3M}\right)^{\frac{1}{3}} R d_{peri}$$

Here, in order to obtain $r_S$ in R(Sun) units, R must be expressed in these units: $R = 4.238 \times m$ .

With the Paczynski-Wiita (1980) approximation for the gravitation field close to the BH, the corresponding expression is

$$(14) \qquad r_S = \left(\frac{m_S}{M}\right)^{\frac{1}{3}} \left(\frac{d_{peri}}{3d_{peri}-1}\right)^{\frac{1}{3}} R(d_{peri}-1)$$

The upper and lower black dashed lines in Figure 5 are the plot of the (13) and (14) mass-radius relations, respectively.

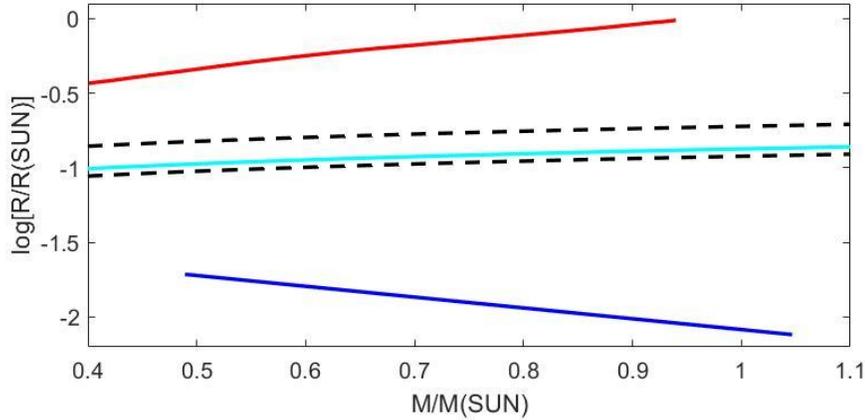

Figure 5: The cyan colored and the two dashed lines represent, on a logarithmic scale, the mass-radius relation of the close companion star of the BH of Sgr A*, as implied by our model, according to the Eggleton (1983) formula and expressions (13) and (14) in this section. The red line is an empirical M-R relation of Galactic low mass stars (Demory et al, 2009). The blue line is the M-R relation of WDs (Romero et al 2019).

The figure demonstrates that the close binary companion star of the BH of Sgr A* must have an unusual internal structure, as will be further discussed in the next section.

## 6. The angular velocity of the star

Figure 6 depicts three presentations of the star polar coordinate $\theta$ as a function of time, on the left-hand side as measured in the S system, and on the right, in the E system. Panels a present this function along 1000 full cycles of $\theta$ that in the E system last 28.25 days. The $\theta(t)$ function is a blue line that is lightly curving above and below a straight red dashed line. The two lines are inseparable by eye in the scale of the two figures. Panels (b1) and (b2) are zooms on a small section of the a plots covering the time of $0.08\ day = 115\ minutes$. Here the two lines are resolved, more clearly on the left-hand side. Panels (c1) and (c2) are half day presentations of the residuals of the blue curves in panels (a) and (b), after removal of the linear trend presented by the red dashed lines.

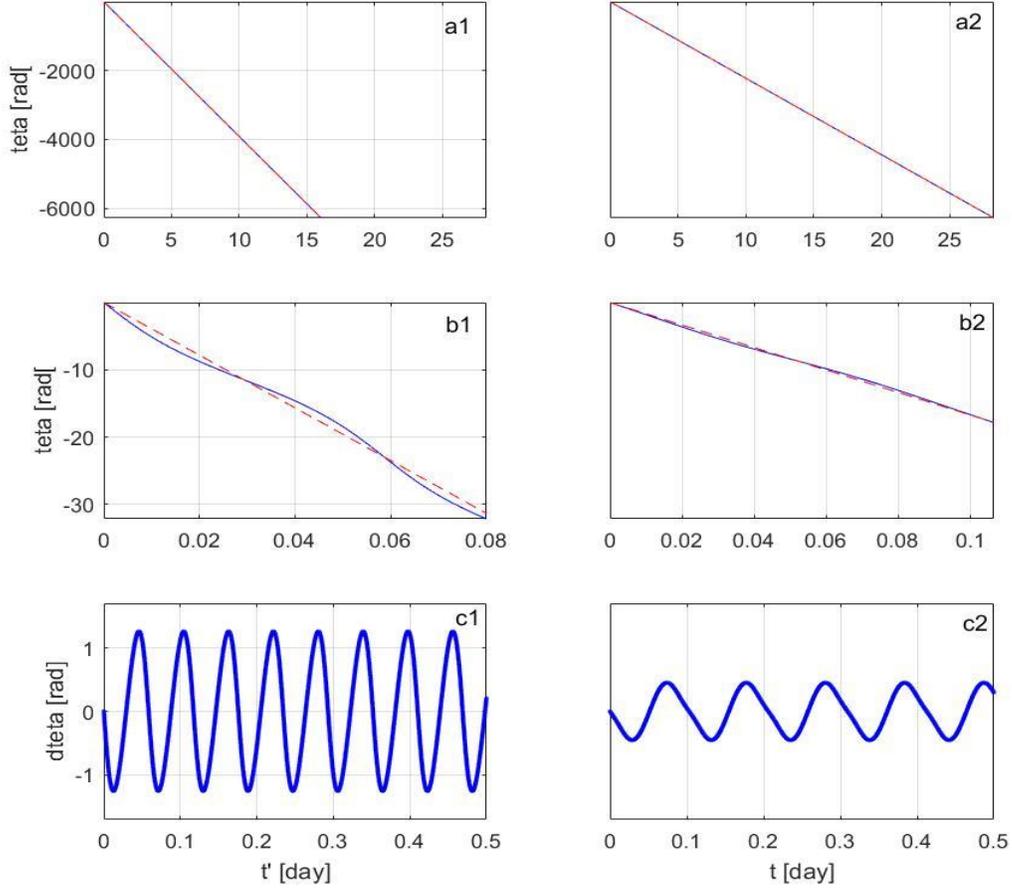

Figure 6: The polar coordinate $\theta$ of the star position on its orbit as a function of time. The two sides of the figure are as in figure 3. Panels (a1) and (a2) present in blue 1000 full $2\pi$ cycles of $\theta$. (b) Zooms on a time interval of 0.08 day of the curves in panels (a1) and (a2), allowing resolving the curving blue line from its linear trend marked by the red dashed line. (c) Residuals of the blue lines after removing the linear red line trend, along one half day time interval.

The slope of the red line in panels (a2) and (b2) is $\omega = -2\pi F_{orb}$. The periodic function in panel (c2) is well represented by the fundamental and the second harmonics of a Sine wave of the $F_{epi}$ frequency. In the left-hand side panels we see the same functional relations between $\theta$, $t'$, $F'_{orb}$ and $F'_{epi}$. Figure 6 exhibits that the angular velocity (AV) of the star in its orbit can be described as periodic oscillations with the epicyclic period of the obit, superposed on a constant angular velocity with the period that is the mean value of periods of many orbital sidereal revolutions of the star.

The most outer layers of the star that are very much deformed by the gravitational tidal forces, and possess only small amount of moment of inertia, are most probably rotating with AV that is locked rigidly to the instantaneous angular velocity of the orbital motion. Along its radius, the star rotation must be highly differential. Presumably up to a certain inner radius $r_c < r_s$ , where $r_s$ is the radius of the star, the rotation is at constant angular velocity, not necessarily but possibly synchronized with $F_{orb}$, the frequency of the mean sidereal orbital AV. At this inner part of the star the self-gravity of the star is stronger than the varying tidal forces that are due to the varying distance of the star from the BH. Up to that radius, the star is mostly spherical and only little torque is exerted on it.

As an example we can imagine a revolving star of mass 1 M(Sun). According to the middle curves in figure 5 its radius is about 0.12 R(Sun). At the star center there is a core of degenerate matter of, say, 0.7 of its mass. In figure 13 of Romero et al (2019) we see that the radius of a WD of this mass is about 0.012 R(Sun). The WD is rotating at a stable sidereal AV. It is engulfed within a thick envelope of mass 0.3 M(Sun) that extends up to the radius of the star of $0.12\, R(Sun)$, occupying some 99.9% of the star volume. This envelope is responding to the varying tidal forces and torques exerted by the BH, as the star revolves around the BH in its highly noncircular orbit described by equation (2). The massive, very thick envelope of the WD is the element in the system that is the major initiator of most of the flaring activity of Sgr A*. A Thorne-Zytkow object (Thorne & Zytkow, 1977) could be a model of a star akin to the one suggested here. Needless to say, however, that a quantitative model describing the structure of a star under the extreme conditions in the close neighborhood of the BH is required in order to substantiate the above suggested explanation of the plots in Figure 5.

## 7. The moving hotpots

In a set of exquisite observations, the GRAVITY Collaboration succeeded in following a motion on the plane of the sky of sources of NIR flares of Sgr A*. For each one of three large NIR flares, recorded on 2018 May 27, July 22, and July 28, they were able to resolve a discrete source, termed hotspot, that was changing its position on the plane of the sky during the lifetime of the flare. For each one of these flares GRAV18 published a map, presenting 10 or 17 different positions of the hotspot as recorded by their extra-ordinary instrument. For the july 22 flare, GRAV18 published a map that is different from the one published in 2020 (Dexter et al. 2020a). I take the later map as the more accurate one and unless stated otherwise, the references to the july 22 map will be to the 2020 version. GRAV18 also published a table of the exact recording times of the 10 spots of the july 22 map. Relying on data contained in figures 2 and B.4 in GRAV18, and on the color convention in Figure 2 of Dexter et al. 2020a, I was able to assign to each spot in the other 2 maps an estimated time of its recording. Even in the cases of the largest uncertainty, the deviation of the assigned time from its true value is probably not larger than 2.5 minutes.

In all maps the spots seem to form some cyclical, though not circular pattern in the plane of the sky. They are all confined to within distances from the median of their distribution with the mean and StD of $7.5 \pm 4$ or $8.5 \pm 6$. Here, and in what follows I utilize GRAV18 relation between angles measured on the plane of the sky and distances at the object plane: $10 \mu as \leftrightarrow 1R$.

### 7.1. The apparent spot motion

What was observed by GRAV18 as motion of spots cannot be the image on the plane of the sky of an emitting mass that moves at the measured rate in the object plane. This is because the rate of change in the position of the centroids of many pairs of successive GRAV hotspots would imply velocities of this mass in the object plane that exceed the speed of light. This and the apparent cyclical nature of the motion lead most researchers to conclude that the mass element at which the flare originates is indeed in a state of some cyclical motion but at small radii so that the rate of change in the position of this mass is subluminal.

One way by which a hotspot appears at a position in the sky that is different from the position of the mass at which it was initiated is a motion of a disturbance away from the position of the mass that gave birth to it. The disturbance travels for some time before it becomes a full-fledged hotspot. Such a mechanism was suggested, for example, by Matsumoto et al (2020) who interpret the GRAV18 map of the july 22 moving spots as exhibiting a pattern motion, possibly excited by a precessing outflow interacting with a surrounding disk. Another approach was adopted by GRAV18. They suggested that the apparent varying positions of the hotspots on the plane of the sky are due to relativistic effects that curve the tracks followed by the photons. The mass emitting these photons is moving on a small circular Keplerian orbit at a velocity that is only a fraction of the speed of light. A combination of these 2 approaches is adopted in other suggested interpretations of the moving hotspots phenomenon. Ball et al (2020) suggested models of plasmoids orbiting in a jet or coronal region of the BH. Effects from finite travel time were included with the general relativistic radiative transfer calculations of photon tracks.

GRAV20 developed a code, more elaborate than the one of GRAV18, to predict astrometric motion as well as flux variability from compact emission regions around the BH. It allows elliptical and out of plane motion and performs ray tracking of photon trajectories. Their model describes spot motion at radii of about 4.5 and motion of the emitting mass at distances from the BH of about 2.5. The paper of Dexter et al (2020b) may be considered as taking one step further by suggesting a detailed mechanism of generating intense energy release events at the rate of the order of the observed rate of NIR flares of the object. A detailed computation is made, describing a development in time of a disturbance originating at very small radii of the inner accretion disk. This disturbance expands and propagates away from its point of origin, leading to the observed scenario of moving centroids of hotspots in the observer plane of the sky.

Linear polarization fraction of 10%-40% is known to characterize the NIR flares radiation (Eckartet al, 2006a; Trippe et al, 2007; Zamaninasab et al, 2010; Shahzamanian et al, 2015). In their spot detection observations GRAV18 found a continuous rotation of the linear polarization angle. The period of the polarization angle rotation matches what is inferred from astrometry.

GRAVITY Collaboration et al (2021, hereafter GRAV21) analyzed in details the magnetic-polarimetric aspects of the moving hotspot phenomenon with a special attention to the july 28 data. They found that with a proper geometry and dynamics of magnetic fields in the zone of the spot generation, the hotspot models of GRAV20a and Dexter et al. (2020b) type can explain the observed evolution of the linear polarization. In the following section I shall deal only with the astrometry of the moving hotspot discovery, assuming, as was found in GRAV21 models, that a successful model of the hotspots within the revolving star scenario will be able to explain also the observed evolution of the linear polarization.

## 7.2. The moving spots in the revolving star model

It seems that most of the above mentioned and other suggested interpretations of the moving hotspot phenomenon, as well as the microphysical, magneto-hydrodynamic processes underlying them (Markoff et al, 2001, Chael et al, 2018, Ripperda et al, 2020, Dexter et al, 2020a and references therein), may be incorporated quite naturally into the model of the close revolving star around the BH. On principle there is only one new element that must be injected into these models and that is where to put in the Sgr A* system the material source of the disturbance that develops into an observed hotspot. I suggest that it is in the atmosphere of the star. From that point on, the timeline of the physical processes that eventually generate the observed spot may be described by some of these other suggested models. This statement should of course be backed by constructing a specific model where the natal place of a NIR Sgr A* flare is in the revolving star. Here I show that such a model is at least feasible.

### 7.2.1. Straight line photon track approximation

In their 2018 paper, GRAV18 presented a spot map of the july 22 event that is, as mentioned above, somewhat different from the one presented in GRAV20. GRAV18 suggested a best-fit circular orbit, on which the position of the source corresponding to each one of the n=10 spots is determined by performing photon track tracing calculations taking into account relativistic effects such as bending, lensing, time dilation and beaming. The best-fit circular orbit that they find has the radius of 3.5.

In this subsection I would like to demonstrate, by way of an example, that at the present accuracy in the measurements of the spot centroid positions, at the typical distances of the recorded spots from the central BH, finding the time of the initiation of a spot by considering a straight line photon track from the center may be a fair approximation of the calculation of a full relativistic curved track.

Following GRAV18 I also looked for circular orbit that best fits the spot map of july 22 as presented in the 2018 paper. The varying parameters are the [x,y] coordinates of the position of the BH, relative to the median position in the GRAV18 map, and the radius a of the assumed circular Keplerian orbit. In order to determine where to put on the orbit the point marking the position of the star when a certain spot was initiated I computed the time of the initiation of that spot by subtracting from its detection time the flight time of a photon along the straight line connecting the BH and the spot. Taking the time of the first spot as 0, this provides nine values $t_j; j = 2:n$ that are the intervals between the time of initiation of the nine later spots and the initiation time of the first one. Each orbit radius a has its Keplerian orbital period $P_{orb}^{.a}$ . The position of a point on the circular orbit is determined by its polar coordinate $\theta$ relative to some reference angle $\theta_{ref}$. Taking $\theta_{ref}$ as defining the location of the star on the orbit at the initiation of the 1st spot, the position on the orbit of the star at the initiation time of the j spot is then given by $\theta_j = \frac{2\pi t_j}{P_{orb}^a}$. I consider $N_c = 1000$ values of the $\theta_{ref}$ angle, covering uniformly the $2\pi$ full cycle. For each [x,y] position of the BH and each radius a of the orbit I can then determine all the possible $N_c$ distributions on the circular orbit of the positions of the star at the initiation moment of the 10 spots.

The best-fit orbit is the one for which the RMS of the distances of the points on the circle to their corresponding spot centroids takes a minimum value. The best-fit circular orbit so found has the same radius 3.5, as found in the GRAV18 analysis. The distribution on the orbital circle of the points that correspond to the 10 spot centroids is also very similar to the distribution of points on the circle shown in Figure 1(d) of GRAV18.

The straight line photon track approximation is not an approximation of the physical track followed by the photons in the environment of the BH. It is used only as a phenomenological-heuristic tool for computing the time delay between the initiation of a disturbance and the appearance of a hotspot in the sky of the observer. The similarity of the outcome of the use of this approximation with the finding of GRAV18 is not very surprising. The geometrical length of the relativistic curved track is most probably somewhat larger than that of a straight line but it is of the same order of magnitude. The difference is partially taken into account by measuring the straight line from the very center and not from a point on the orbit. The time delay is then depending on the velocity of the photon which is the same velocity of light  c  on all tracks.

### 7.2.2. The three GRAV20 spots maps and the revolving star

In the revolving star scenario, for the assumed m=4.2184, the orbit of the star is given by equation (2). The location of the star on its orbit at any time t measured from $t_0$, can be obtained from equation (3). For each map, using the straight line photon track approximation, I find the n-1 time intervals between the moments of the initiation of the n-1 later spots and that of the first one. The conversion of the times to positions on the orbit is performed with the help of equation (3). Here, the orbit itself is given. The search is then for a section of the 28 day long orbit that best fit the spots in the sense defined in the previous section. This is done by trying each one of nine locations around the median point of the map as the position of the BH, and by scanning over the $10^6$ calculated points of the orbit, considering each one of them as the position of the star on the orbit at the moment of initiation of the first spot. This determines the positions of the star at the other nine moments of initiation.

Figure 7 presents the results of the best-fit procedure applied on the 3 maps displayed in Figure 2 in Dexter et al. (2020b). The star orbit is shown with an inclination angle  $i = 0^0$  of the orbital plane with respect to the plane of the sky. Similar results and images are obtained with all $i \leq 20^0$ values. The scale of the [x,y] axes is in R units. Large black crosses are the observed spot centroid positions and the error bars in GRAV20a maps. The one with the green point marks the first spot that was recorded in each observing night. The rotations of the spots and of the star are in the clockwise direction. The small magenta cross is the median position in the corresponding GRAVITY map. "orbit" in the figures are the count numbers in sidereal binary revolutions following the initial condition, of the beginning and end points of the displayed orbital section. <R> is the mean orbital radius of the displayed section.

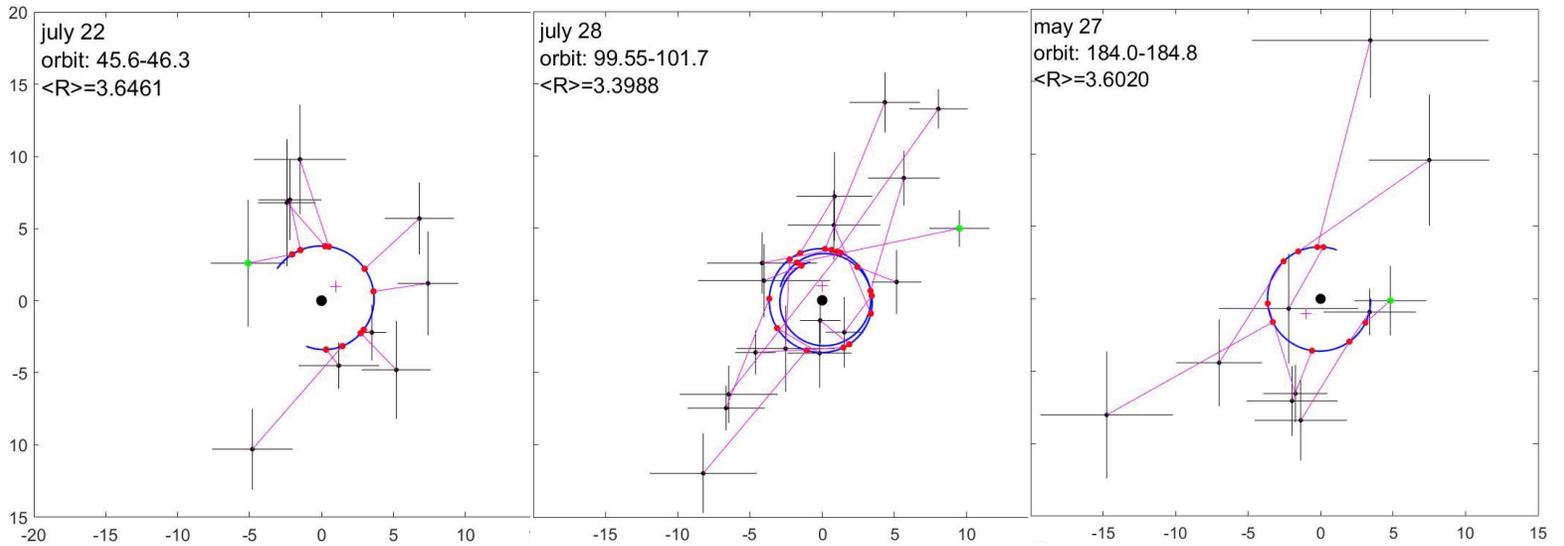

Figure 7: Maps of the hotspots associated with three large NIR flares of Sgr A* recorded in three nights by the GRAVITY collaboration. The x and y scales are in $R_{Schwarz.}$ units. The cross with the green point at the center is the one recorded first. The small magenta cross is the position of the median point in GRAV20a maps. The blue curve is the section of the star orbit presented by equation (2) that best fit the spot positions and the sequence of their recording times in each night. The parameters "orbit" give the sidereal binary cycle numbers of the beginning and end points of the displayed orbital section, <R> is the mean radius of the displayed section. See text for explanations.

There is some noticeable similarity between the left and the right panels of Figure 7 and the corresponding maps of the flares of July 22 and May 27 in Figure 2 in GRAV20. For the July 28 flare, the distribution of the point marking the position of the star on the suggested orbit is quite different from the one suggested by GRAV20. In particular, the best fit process applied here suggests that the long flare of that date lasted over 2 full cycles of the sidereal binary revolution of the star while in GRAV20 the flare took place along only one such revolution. Some possible support for the two cycles proposition comes from Figure 5 of GRAV20 paper. There it is shown that the LC of the flare of July 28 has a double hump structure. This may be a reflection of the expected variation of the measured intensity of a flare that lasts over 2 cycles of the sidereal binary revolution of the star rather than over just 1, as discussed above in Section 4.7.

### 7.2.3. A different approach

Dexter et al. (2020b) presented detailed calculations of a model whereby magnetically dominated plasma is expelled from near the BH horizon and forms a rotating spiral structure. The increase in the NIR emission that creates a hotspot in the disk, occurs via reconnection at the interface of the magnetically dominated plasma and surrounding fluid.

I suggest that the initial matter and energy injection that is the seed of the NIR flare phenomenon is in the atmosphere of the star. It is a stellar flare-like event. The injections of the magnetic dominated plasma may be up scaled solar type events such as magnetic flux tubes, closed or open, that protrude into the surrounding medium, or a jet or mass ejection, akin to the solar coronal mass ejection events, that are thrown into it. As in Dexter et al. (2020b) model, the injection of this mass and magnetic energy is a seed of the formation in the disk of the zone of high density, temperature and fluid heating rate. The internal dynamics within the hot zone that gives it its spiral structured is determined primarily by the circular velocity of the accreted matter of the disk. Along the duration of a flare, the entire spiral pattern is rotating with the rotation of its seed, namely, at the orbital AV of the star. The idea that stellar winds are important elements in the process that generates the variable NIR radiation of Sgr A* was raised also by Ressler et al. (2020) although in a context of an entirely different scenario.

In a model that combines the revolving star scenario with models of the type suggested by Dexter et al. (2020b) and GRAV21, we are able to determine as before the positions of the star on its orbit at the time of the spot initiations. For this we have to make either one of two further assumptions regarding the physical conditions at the site of the hotspot creation. Admittedly, so far there is no observational or theoretical foundation for making these assumptions. It should be noted, however, that neither one of these assumptions is a necessary condition for considering such a combined model a feasible one, although without any one of them it would be difficult or impossible to determine the positions of the star at the initiation moments.

We may assume that in each flare event there is a constant time interval $\Delta T$ between the germination of the seed of a hotspot at the atmosphere of the star and its materialization as the discernible hotspot on the surface of the disk by an observer on Earth. Under this assumption the time intervals between positions of the star on its orbit are equal to the intervals between the detection times of the corresponding spots. The maps that are obtained in this way are naturally not very different from those displayed in figure 5, since the orbit considered is the same one expressed by equation (2). However, the orbit sections that are found best fitted to the maps are in this case different from those displayed in figure 7.

Alternatively, we may assume that the velocity u of a disturbance that propagates in the plasma medium of the disk, from the position of the star at the moment of its initiation to the position of the detected centroid of the corresponding spot, is the same in all directions. Under such assumption the best fit orbital section is found as in section 6.2.2, except that here the distance of each spot from the BH is divided by the velocity u, rather than by c, as done in that section. We find that in this case, in order to obtain a reasonable fit, u must be a highly relativistic velocity, $u > 0.8c$.

## 8. Summary

This paper presents some observational evidence for the operation in the Sgr A* system of a third pacemaker with a period $P_{pre} \cong 56\ min$ that regulates the statistics of the recorded times of midpoints of NIR flares of this object. This is in addition to the NIR pacemaker with the period $P_{orb} = 41\ minutes$ which was reported about in L3. The other pacemaker that operates in the system has the period $P_{epi} = 149\ minutes$, regulating the statistics of the X-ray flares of the object. As an interpretation of these findings I suggested in L3 a model in which a star is orbiting the BH in a quasi-Keplerian precessing elliptical orbit with a semi-major axis, here slightly corrected, a=3.1278. In this paper I presented a detailed orbit along which the star may be moving in its revolution around the BH. The agents in the Sgr A* system that are responsible for the three pacemakers can be recognized in the time series of three kinematical parameters of this orbit.

The star is found to have a peculiar internal structure, not obeying the mass-radius relations that characterize neither common low mass stars nor WDs. I hypothesized that the star has a very dense core, probably consisting of degenerate matter. The core rotates with a constant AV, possibly synchronized with the angular velocity of a circular Keplerian revolution at a distance $a(1 - e^2)$ from the BH. The outer layers of the star respond vehemently to the violently varying gravitational force in the spacetime in which the star is moving. The rotation AV of the most outer ones is probably locked rigidly to the instantaneous AV velocity of the star in its orbit suggested here. Naturally, all these hypothetical qualities of the star must be backed by a consistent quantitative model for such a star.

I suggest that the very origin of the energy release that brings about the appearance of the moving hotspots that were detected by the GRAVITY Collaboration is in this revolving star. It is shown that an interpretation of the moving hotspots phenomenon on the basis of such a model yields results that fit the observed data in a similar qualitative manner as those obtained by other models that have been suggested as interpretation of this phenomenon. In fact, there is no real conflict between the model suggested here and some of the other models. The difference is mainly where is the point of origin of a detected hotspot centroid of an NIR flare of Sgr A*. In the model presented here it is in the revolving star.

There are three observational facts that so far have not been addressed by other models of Sgr A* flaring phenomenon, that gain a plausible explanation in the revolving star model.
1. The finding of the values $P_X$, $P_{IR}$ and $P_{IRM}$ as the periods of the pacemakers that are associated with of the three sets of detection times of Sgr A* flares that have small FNPs to be considered accidental.
2. The finding that each X-ray flare is followed by an NIR flare (Boyce et al. 2019). How this coincidence is finding an explanation within the revolving star model is explained in Section 8.3.2 in L3.
3. The time scale of the apparent cyclic motion of the spots in all three GRAV (Dexter et al. 2020b) NIR flares that is of the order of the period of a Keplerian orbit with a major axis $a \approx 3.1$.

If the revolving star model is indeed a viable presentation of an astronomical reality it may be used also for tracing possible detectable evolution of the very close binary BH-star system that might take place on human lifetime scale. For this purpose, more measurements of the precise timing of Sgr A* flares are required, in order to extend the time basis of the recorded events. Additional observations may of course also serve as further tests of the quality of the statistical analysis performed here and in the other papers in this series. The model itself will probably have to be constructed on firmer grounds of more accurate, fully relativistic equations that govern the dynamics of the system. On the basis of better data and theoretical foundation, signs of such an evolution may perhaps be detected, or at least some constraining limits on the rate of change in this dynamical system may be established.

The model relies on two supporting legs. Observationally, its necessity and validity depend on the statistical significance that one can assign to the discovery in the observed data the particular values of the periods of the three pacemakers associated with the observed data sets. If these are eventually found to be statistical flukes, the model loses its main observational support and motivation. On the theoretical side, if it can be shown that the orbit of the star as suggested in this paper is dynamically unfeasible, or that it cannot be stable for more than a very few years, the model should obviously be abandoned. In that case, however, if the statistical confidence that the period of at least one of the three pacemakers that are associated with the three time sets of Sgr A* flares is not a statistical noise remains un-shattered, or even strengthens in the future, any future model for this object flaring phenomenon will have to provide an explanation of this observational finding.

## Acknowledgment

I thank an anonymous referee for some critical comments that led to considerable improvement in the analysis and in the quality of the results presented in this paper.